\documentclass[journal]{IEEEtran}
\usepackage{amsmath,amsfonts}
\usepackage{algorithmic}
\usepackage{array}
\usepackage{textcomp}
\usepackage{stfloats}
\usepackage{url}
\usepackage{hyperref}
\hypersetup{hidelinks}
\usepackage{verbatim}
\usepackage{graphicx}
\def\BibTeX{{\rm B\kern-.05em{\sc i\kern-.025em b}\kern-.08em
    T\kern-.1667em\lower.7ex\hbox{E}\kern-.125emX}}
\usepackage{balance}

\usepackage{subcaption}
\usepackage{xfrac}
\usepackage[dvipsnames,table]{xcolor}
\usepackage{bm}
\allowdisplaybreaks
\usepackage{cuted}
\usepackage{ctable}
\usepackage{siunitx,booktabs}
\usepackage[bottom]{footmisc}

\usepackage[margin=0.60in]{geometry}

\pdfoutput=1

\begin{document}
\title{Inertia-Aware Microgrid Investment Planning Using Tractable Decomposition Algorithms}

\author{Agnes~Marjorie~Nakiganda,~\IEEEmembership{Member,~IEEE,
}
Shahab~Dehghan,~\IEEEmembership{Senior~Member,~IEEE,}
and~Petros~Aristidou,~\IEEEmembership{Senior~Member,~IEEE}
        
\thanks{A. Nakiganda is with the Department of Electrical Engineering, Technical University of Denmark, Denmark (email: amanak@dtu.dk). 

S. Dehghan is with the School of Engineering, Newcastle University, UK.

P. Aristidou is with the Department of Electrical Engineering and Computer Engineering and Informatics, Cyprus University of Technology, Cyprus.\vspace{0.5em}}
}
\maketitle

\begin{abstract}
    The integration of the frequency dynamics into Micro-Grid (MG) investment and operational planning problems is vital in ensuring the security of the system in the post-contingency states. However, the task of including transient security constraints in MG planning problems is non-trivial. This is due to the highly non-linear and non-convex nature of the analytical closed form of the frequency metrics (e.g., frequency nadir) and power flow constraints. To handle this issue, this paper presents two algorithms for decomposing the MG investment planning problem into multiple levels to enhance computational tractability and optimality. Furthermore, the sensitivity of the decisions made at each level is captured by corresponding dual cutting planes to model feasible secure regions. This, in turn, ensures both the optimal determination and placement of inertia services and accelerates the convergence of the proposed decomposition algorithms. The efficient and effective performance of the proposed algorithms is tested and verified on an 18-bus Low Voltage (LV) network and a 30-bus Medium Voltage (MV) network under various operating scenarios.
\end{abstract}


\section{Introduction}
\IEEEPARstart{T}{he} ability to operate in dual modes (i.e., connected to the grid or disconnected from the grid as autonomous islands) has cemented Micro-Grids (MGs) as pivotal infrastructure to increase the resilience of the grid. In other words, MGs are capable of enhancing the resilience of the system by forming autonomous islands during extreme events in the grid, and consequently, ensuring the continuity of the power supply \cite{7091066,izadi2021critical}. However, this is possible only if security is ensured {\it before}, {\it during}, and {\it after} MG islanding events. Therefore, security analysis needs to include the evaluation of both {\it pre-} and {\it post-islanding} operating points, as well as the system trajectory during the transition, to ensure the operation is within acceptable limits and secure regions. Hence, the integration of the frequency dynamic evolution {\it during} islanding into MG investment and operational planning problems is vital for improving the security of the system in the post-contingency states, as well as system preparedness. 

While steady-state security has been widely studied in MG planning problems using the power flow model and associated static constraints \cite{6509996,8765789,6671442,8756210}, dynamic security requires the inclusion of the system transient evolution model. The latter is usually represented by Differential Algebraic Equations (DAEs) and associated time-domain trajectory constraints defined by Grid Code regulations. However, adding dynamic security constraints in MG planning problems results in a highly complex optimization problem, which cannot be easily solved by off-the-shelf optimization packages. 

A more tractable alternative strategy to evaluate dynamic security is with the use of analytically derived expressions for transient frequency evolution \cite{Anderson1990,kundur1994power,UrosLQR}, that can be embedded into MG planning problems as proxies for trajectories of DAEs to assess different metrics ensuring the security of the system. These security metrics consist of frequency nadir, frequency zenith, Rate-of-Change of Frequency (RoCoF), and the steady-state post-transient frequency deviation. Note that these analytical expressions map the ``control parameters'' (i.e., system inertia, damping, and droop levels) to ``system states'' essential during frequency support. 
Although including analytical expressions in MG planning problems to evaluate the transient frequency evolution is more tractable than including DAEs, their highly non-linear characteristics result in a complicated optimization problem. 

The time-domain trajectory of the system frequency deviation has been described by a first-order ordinary differential equation in \cite{kundur1994power}. This dynamic model has been adopted in \cite{7115982} to derive expressions for the frequency nadir, RoCoF, and quasi-steady-state frequency imposed on a stochastic scheduling problem. 
It is noteworthy to mention that the analytical expressions for the transient frequency security metrics have been previously derived in \cite{Anderson1990} considering only Synchronous Generation (SG) units and recently enhanced in \cite{UrosLQR} to include the support from Converter Interfaced Generation (CIG) units. While the RoCoF and quasi-steady-state constraints are convex, the non-linear frequency nadir expression in \cite{UrosLQR} has been approximated by a bi-linear constraint with further sufficient conditions on the approximation defined by mixed-integer linear constraints. Sufficient conditions derived from a pre-determined frequency trajectory have been applied to the linearized frequency nadir constraints in \cite{7310887} for unit commitment and economic dispatch of power generators. In \cite{8667397}, a two-step linearization technique defined by an inner approximation utilizing overestimating planes and the standard big-M technique has been introduced to linearize the nadir expression. Furthermore, the work in \cite{9099053} has transformed the frequency nadir constraints into the capacity reserve constraints using a series of linear frequency security margin constraints formulated by piece-wise linearization fitting of the non-linear expression. Similarly, \cite{6717054} has tackled this problem by approximating the nadir expression using a piece-wise linearization technique. In the linearized form, the expression has been suitably integrated into a security-constrained unit commitment problem with frequency limits applied. Additionally, the work in \cite{UrosUC} has utilized ex-ante bound extractions on the variables of the nadir expression using potential dispatch conditions that have been then imposed to the security-constrained unit commitment problem replacing the non-linear nadir expression. Furthermore, the non-linear frequency nadir expression in \cite{9165193} has been characterized and approximated by using a neural network.

The aforementioned studies have certain inaccuracies emanating from the adoption simplified dynamic models, approximations of the frequency nadir constraint or ex-ante limit extractions to simplify the planning model. Moreover, the characteristic properties of MGs have not been included in these studies. Due to the inaccuracy inherited from different approximations, a three-stage iterative algorithm with a sequential linearization has been proposed in \cite{9691464} using the more accurate low-order frequency model in \cite{UrosLQR}. The first stage of the problem ensures the optimality of the planning solution, while the second stage is a feasibility checking problem against the transient security problem. In the third stage, tighter bounds for power exchange with the main grid are formulated and generated for the first stage problem if infeasibilities exist in the second stage. Although the three-stage algorithm in \cite{9691464} does not require any approximation of transient frequency constraints when applied to the optimization problem, it finds a rather conservative solution. Additionally and more importantly, similar to the previous literature, the techno-economic effect of different generator parameters on the planning solution and security limits has not been adequately studied in \cite{9691464}.  

Motivated by the challenges mentioned above, two efficient multi-level decomposition strategies, utilizing dual cutting planes based on the notion of Benders decomposition, are introduced in this paper to more cost-effectively solve the model presented in \cite{9691464}. The dual cutting planes in the proposed decomposition strategies ensure that the techno-economic effects of the MG's parameter changes in the {\it feasibility-checking} problem are cost-effectively captured by the {\it investment and operational planning} problem. The proposed approach, therefore, allows for the decoupling of frequency services where emphasis can be applied to single or multiple services provision. This ensures optimal determination and provision of frequency services in MGs. Accordingly, the main contributions of this paper are threefold:


\begin{itemize}
    \item Two multi-level decomposition strategies are proposed to ensure tractable incorporation of non-linear transient frequency security constraints in the MG investment and operational planning problem. The proposed decomposition strategies utilize dual cutting planes to capture the effect of decisions made in the feasibility-checking sub-problems on decisions made in the investment and operational planning master problem. Subsequently, this enables the segmentation of different frequency services that can be provided by each generator or the system as a whole.
    \item To linearize the non-linear frequency nadir constraint, a first-order Taylor expansion is adopted in this paper. Note that higher orders can be utilized to enhance the accuracy of the solution at the expense of significantly higher computational time.
    \item The functionality of the proposed decomposition algorithms is compared against the bound-tightening approach proposed by the authors in \cite{9691464}, under different operational scenarios on the 18-bus CIGRE European Low-Voltage (LV) network and a 30-bus Medium-Voltage (MV) distribution network. Case studies with time-domain simulations demonstrate the superiority of the proposed decomposition strategies over the three-stage solution approach presented by the authors in \cite{9691464}. 
    
\end{itemize}

The remainder of this paper is organized as follows. Section \ref{prelim} presents the frequency response algorithm and the formulation of the planning problem. Section \ref{model} describes the proposed decomposition strategies for the inertia-aware planning problem. Then, the case study results are presented in Section \ref{study}. Finally, the main conclusions are drawn in Section \ref{concl}.

\section{Preliminaries}\label{prelim}
\subsection{Frequency Response Model}

The frequency response model utilized here is based on the uniform representation of frequency transients primarily introduced in~\cite{Anderson1990} for a traditional system with only SG units and extended in~\cite{UrosLQR} for a low-inertia system with both SG and CIG units. It is worth mentioning that different generation units of the system may have moderately different transient frequency response after a disruptive event. However, the dynamic performance characterized by the Center of Inertia (CoI) swing equation has been verified to adequately capture the transient performance of the system~\cite{Anderson1990,UrosLQR}. In~\cite{UrosLQR}, the dynamic model includes SG units (indexed by $i \in \mathcal{S}$) and CIG units (indexed by $c \in \mathcal{C}$). Furthermore, the characteristics of {\it grid-supporting} CIG units providing frequency support through droop (indexed by $d\in \mathcal{C}^d\subseteq\mathcal{C}$) and Virtual Synchronous Machine (VSM) (indexed by $v \in \mathcal{C}^v\subseteq\mathcal{C}$) control strategies, as the most popular control techniques in the literature ~\cite{Rocabert2012,UrosGM}, are embedded in the frequency response model. The transfer function $G(s)$ between the active power change $\Delta P_e(s)$ and the CoI frequency deviation $\Delta f(s)$ can be derived as given below:   
\begin{align}\label{eq:G1}
\begin{split}
    G(s) &= \dfrac{\Delta f(s)}{\Delta P_e(s)} = \Biggl(\overbrace{ \vphantom{\sum\limits_{i\in\mathcal{S}} \dfrac{K_{i} (1+sF_{i} T_{i} )}{R_{i}(1+sT_{i})}}(sM_s+D_s )}^{\text{SGs Swing Dynamics }} +\overbrace{\sum\limits_{i\in\mathcal{S}} \dfrac{K_{i} (1+sF_{i} T_{i} )}{R_{i}(1+sT_{i})}}^{\text{SGs Turbine \& Governor Response}}\\
    & \hspace{1cm}+ \overbrace{\sum\limits_{d \in \mathcal{C}^d} \dfrac{K_{d}}{R_{d} (1+sT_{d} )}}^{\text{Droop-based CIGs}} + \overbrace{\sum\limits_{v\in\mathcal{C}^v } \dfrac{sM_{v}+D_{v}} {1+sT_{v}} }^{\text{VSM-based CIGs}}\Biggr)^{-1}. 
\end{split}
\end{align}

\noindent
where, a positive value of the active power change $\Delta P_e(s)$ corresponds to a net load decrease. Moreover, $M_s$ and $D_s$ represent the aggregated normalized inertia and damping of all SG units, respectively, while $K_i$, $F_i$, $T_i$, and $R_i$ denote the mechanical gain factor, fraction of total power generated by the SG turbine, time constant of the SG turbine, and droop of the SG unit $i$, respectively. In addition, $K_{d}$, $R_{d}$, and $T_{d}$ represent the power gain factor, droop, and time constant of the droop-based CIG unit $d$, respectively, while $M_{v}$, $D_{v}$, and $T_{v}$ denote the virtual inertia constant, virtual damping constant, and time constant of of the VSM-based CIG unit $v$. It is worth noting that CIG units with a droop control strategy incorporate only the damping ability of the converter (i.e., $D_d = 1/R_d$) while CIG units with a VSM control strategy incorporate both the damping and inertia abilities of the converter (i.e., $D_v$ and $M_v$, respectively)~\cite{UrosLQR}. 

It is worth mentioning that local frequency oscillations result in a distinct transient response at each generator after a disruptive event. However, the dynamic performance characterized by the CoI swing equation using aggregated normalized inertia $M$ and damping $D$ has been shown to provide a smooth overall frequency with investigations showing adequate capture of the transient performance of the system~\cite{Anderson1990,UrosLQR,8260199,8450880}.
Moreover, the CoI frequency facilitates the provision
of better control from CIGs~\cite{8260199,8450880}.

Since the time constants of all SG units $(T_i\approx T)$ are orders of magnitude higher than those of CIG units, it can be assumed that $T \gg T_{d,v}\approx0$ ~\cite{UrosStab}. Therefore, the transfer function $G(s)$ in \eqref{eq:G1} can be approximated as:
\begin{equation}
    G(s) = \frac{1}{MT}\frac{1+sT}{s^2+2\zeta\omega_n s + \omega_n^2}, \label{eq:G2}
\end{equation}
where, $\omega_n = \sqrt{\frac{D+R_s}{MT}}$ and $\zeta = \frac{M+T(D+F_s)}{2\sqrt{MT(D+R_s)}}$. Also, other parameters in \eqref{eq:G2} can be calculated as follows:

\begin{subequations}\label{eq:fre-met param}
\begin{align}
&M_s = \sum_{i\in\mathcal{S}} M_{i}\dfrac{P_{i}}{P_{s}^{\mathrm{base}}}, \quad\,\, D_s = \sum_{i\in\mathcal{S}} D_{i}\dfrac{ P_{i}}{P_{s}^{\mathrm{base}}},\\
&R_s = \sum_{i\in\mathcal{S}} \dfrac{K_{i}}{R_{i}}\dfrac{P_{i}}{P_{s}^{\mathrm{base}}},\quad\,\,\, F_s = \sum_{i\in\mathcal{S}} \dfrac{K_{i}F_{i}}{R_{i}}\dfrac{P_{i}}{P_{s}^{\mathrm{base}}},\\
&M_c = \sum_{v\in\mathcal{C}^v} M_{v}\dfrac{P_{c_v}}{P_{c}^{\mathrm{base}}}, \;\;\, D_c = \sum_{v\in\mathcal{C}^v} D_{v}\dfrac{ P_{c_v}}{P_{c}^{\mathrm{base}}},\\
&R_c = \sum_{d \in \mathcal{C}^d}R_{d} \dfrac{P_{c_d}}{P_{c}^{\mathrm{base}}},\\
&M = \dfrac{M_sP_{s}^{\mathrm{base}} + M_c P_{c}^{\mathrm{base}}}{P_{s}^{\mathrm{base}} +P_{c}^{\mathrm{base}}}, \label{eqn: Msup}\\
&D = \dfrac{D_sP_{s}^{\mathrm{base}} + D_c P_{c}^{\mathrm{base}}+ R_c P_{c}^{\mathrm{base}}}{P_{s}^{\mathrm{base}} +P_{c}^{\mathrm{base}}}, \label{eqn: Dsup}
\end{align}
\end{subequations}
where, $P_i$ and $P_c$ denote the active power capacity of the SG unit $i$ and the CIG unit $c$, respectively. Furthermore, $P_{s}^{\mathrm{base}}$ and $P_{c}^{\mathrm{base}}$ represent the base power of all SG and CIG units connected to the system, that is, $P_{s}^{\mathrm{base}} =  \sum_{i\in\mathcal{S}}  P_{i}$ and $P_{c}^{\mathrm{base}} =  \sum_{c\in\mathcal{C}}  P_{c}$, respectively.

The dynamic frequency response of the system after a disturbance can be represented by frequency nadir (i.e., $\Delta f^\mathrm{max}$) and instantaneous RoCoF (i.e., $\dot{f}^\mathrm{max}$) while the steady-state frequency response of the system can be described by the constant frequency deviation from a pre-disturbance equilibrium (i.e., $\Delta f^\mathrm{qss}$). Given a step-wise disturbance in the active power $\Delta P_e(s) = -\Delta P/s$, where $\Delta P$ denotes the net change in the active power, the mathematical expression of the frequency deviation in the time domain (i.e., $ \Delta \omega(t)\equiv \Delta f(t) $) can be obtained as follows:
\begin{equation}\label{model3_coef}
    \resizebox{1\hsize}{!}{%
        $\Delta \omega(t) = -\dfrac{\Delta P}{M}\Bigg(\dfrac{1}{T\omega_n^2} + \dfrac{1}{\omega_d}e^{-\zeta\omega_n t}\left(\sin{\omega_d t}-\dfrac{1}{\omega_n t}\sin{\omega_d t + \phi}\right)\Bigg),$%
        }
\end{equation}
where, $\omega_d = \omega_n\sqrt{1-\zeta^2}$ and $\phi = \sin^{-1}\left(\sqrt{1-\zeta^2}\right)$. 

The frequency nadir occurs at the time instance $t_m$ when $\Delta \Dot{\omega}(t_m) \equiv \Delta \dot f(t_m) = 0$. Therefore, $t_m$ can be derived by solving $\Delta \Dot{\omega}(t_m) \equiv \Delta \Dot{f}(t_m) = 0$, and then the frequency nadir can be obtained as given below:
\begin{equation}\label{eq:nadir}
\resizebox{1\hsize}{!}{%
    $\Delta f^\mathrm{max} = \Delta f(t_m) = - \frac{\Delta P}{D+R_s} \left( 1 + \sqrt{\dfrac{T(R_s-F_s)}{M}} e^{-\zeta\omega_n t_m}\right),$}
\end{equation}
where, $t_m =(\sfrac{1}{\omega_d})\tan^{-1}\left(\sfrac{\omega_d}{\left(\omega_n\zeta - T^{-1}\right)}\right)$.

In addition, the maximum RoCoF occurs at the instance of the disturbance (i.e., $t_r = 0^+$). Therefore, the maximum RoCoF can be obtained by solving $\Delta \Dot{\omega}(0^+) \equiv \Delta \Dot{f}(0^+)$ as follows:
\begin{align}
    \Delta \dot{f}^\mathrm{max} &= \Delta \dot{f}(0^+) = -\frac{\Delta P}{M}, \label{eq:rocof}
\end{align}

Finally, the quasi-steady-state frequency deviation can be derived from $\Delta \omega (t) \equiv \Delta f(t)$ in \eqref{model3_coef} for $t\to \infty$ as given below:
\begin{align}
    \Delta f^\mathrm{qss} &= \Delta f(\infty)=  -\frac{\Delta P}{D + R_s}. \label{eq:qss}
\end{align}

It can now be concluded that the system control parameters $M$, $D$, $R_s$ and $F_s$ can directly affect the frequency performance of the system. In other words, while frequency nadir in \eqref{eq:nadir} is a highly non-linear function of all four system control parameters mentioned above, RoCoF and quasi-steady-state frequency deviation are explicitly affected by $M$ in \eqref{eq:rocof} and both $D$ and $R_s$ in \eqref{eq:qss}, respectively. Therefore, the above frequency security metrics can be integrated into the MG planning problem to determine cost-effective generation technologies with optimal control parameters to ensure system security in the event of a severe disturbance. 
\subsection{Formulation of Inertia-Aware Expansion Planning Problem}
Similar to the previous work of the authors in \cite{9691464}, a single-year mathematical formulation is considered in this paper to solve the inertia-aware investment and operational planning problem for MGs. In the proposed model, uncertain variations of load demands and renewable generations during the planning horizon are characterized by a sufficient number of representative days, obtained by utilizing the $k$-means clustering technique \cite{dehghan2019robust}. Also, it is assumed islanding from the main grid may occur at each hour (indexed by $t \in \mathcal{T}$) of every representative day (indexed by $o \in \mathcal{O}$) to ensure the robustness of the optimal investment and operational plan against the worst-case unscheduled event in MG. Therefore, the compact formulation of the proposed planning model can be presented as follows:       
\begin{subequations}
\label{eqn:compact2}
\begin{alignat}{3}
     & \min_{\forall \bm \chi}  \Theta^{{\mathrm{inv}}}(\bm \chi^{\mathrm{inv}})+\Theta^{\mathrm{gm,opr}}(\bm \chi^{\mathrm{inv}},\bm \chi^{\mathrm{gm,opr}}) +\gamma  \hspace{-1.4cm} \label{eqn:com11}\\
    & \textrm{subject to:} \nonumber & \\
    &   \gamma\geq\Theta_{to}^{\mathrm{im,opr}}(\bm \chi^{\mathrm{inv}},\bm \chi^{\mathrm{gm,opr}},\bm \chi^{\mathrm{im,opr}}), &  \forall t\in\mathcal{T},o\in\mathcal{O}, \label{eqn:com 21}\\
    & \Phi^{\mathrm{gm,opr}}(\bm \chi^{\mathrm{inv}},\bm \chi^{\mathrm{gm,opr}}) \leq 0, \label{eqn:inequal-gm}\\
     & \Phi^{\mathrm{im,opr}}(\bm \chi^{\mathrm{inv}},\bm \chi^{\mathrm{gm,opr}},\bm \chi^{\mathrm{im,opr}}) \leq 0, \label{eqn:inequal-im}\\
    &  \underline{\Delta f}^\mathrm{max} \leq \Delta f^\mathrm{max}_{to} \leq \overline{\Delta f}^\mathrm{max},\; &\forall t\in\mathcal{T},o\in\mathcal{O}, \label{eq:nadir-const}\\
    & \underline{\dot{f}}^\mathrm{max} \leq \dot{f}^\mathrm{max}_{to} \leq \overline{\dot{f}}^\mathrm{max},\; &\forall t\in\mathcal{T},o\in\mathcal{O}, \label{eq:rocof-const}\\
    & \underline{\Delta f}^\mathrm{qss} \leq \Delta f^\mathrm{qss}_{to} \leq \overline{\Delta f}^\mathrm{qss},\; &\forall t\in\mathcal{T},o\in\mathcal{O}, \label{eq:qss-const}
\end{alignat}
\end{subequations}
where scalar variables are indicated by non-bold symbols while vectors/metrics are indicated by bold symbols. Also, $\bm \chi$ denotes the vector of all decision variables relating to the investment $\bm \chi^\mathrm{inv}$, grid-connected operation $\bm \chi^\mathrm{gm,opr}$, and islanded operation $\bm \chi^\mathrm{im,opr}$ of the MG, i.e., $\bm \chi = [\bm \chi^\mathrm{inv}, \bm \chi^\mathrm{gm,opr},\bm \chi^\mathrm{im,opr}]$. 

The objective function \eqref{eqn:com11} minimizes the total investment costs (i.e., $\Theta^{{\mathrm{inv}}}$), the {\it ``expected''} total operational costs in the grid-connected mode (i.e., $\Theta^{{\mathrm{gm,opr}}}$) for all hours of all representative days, and the {\it ``worst-case''} total penalty costs of disconnecting load demands from MG in the islanded mode for the worst operational hour among all hours of all representative days (i.e., $\Theta^{{\mathrm{im,opr}}} \equiv \gamma$). In this paper, the auxiliary variable $\gamma$ is used to minimize the penalty costs for the worst operational hour among all hours of all representative days in the islanded mode  (i.e., $\Theta^{{\mathrm{im,opr}}} \equiv \gamma \geq \Theta^{{\mathrm{im,opr}}}_{t,o} \quad \forall t \in \mathcal{T},\forall o \in \mathcal{O}$) rather than the aggregated penalty costs for all hours of all representative days.

Constraints \eqref{eqn:inequal-gm} and \eqref{eqn:inequal-im} correspond to the static linearized operational limitations in the grid-connected and islanded modes, respectively. Note that equality constraints (i.e., $a = b$) can be included in \eqref{eqn:inequal-gm} and \eqref{eqn:inequal-im} by opposite inequality constraints (i.e., $a \geq b$ and $a \leq b$). Also, constraints \eqref{eq:nadir-const}, \eqref{eq:rocof-const}, and \eqref{eq:qss-const} represent the transient frequency security limitations to ensure the adequacy of the operational reserve for frequency support in the event of islanding from the main grid, where $\underline{\bullet}/\overline{\bullet}$ denotes the lower/upper limitation of the quantity $\bullet$. In this paper, it is assumed that the transient frequency response, as described in \eqref{eq:nadir}, \eqref{eq:rocof}, and \eqref{eq:qss} and utilized in \eqref{eq:nadir-const}, \eqref{eq:rocof-const}, and \eqref{eq:qss-const}, respectively, depends on the amount of power exchange with the main grid at the time of islanding of the MG, i.e., $-\Delta P = p^{\mathrm{grid}}_{to}$. Therefore, the frequency support provided by a generator $g$, given its investment status $z_g$, is a function of its control parameters, i.e., $M(z_{g})$, $D(z_{g})$, $F_g(z_{g})$, and $R_g(z_{g})$ in addition to the amount of power exchange with the main grid at the time of islanding.  The extended formulation of the investment planning model utilized in this paper can be found in \cite{9691464}. 

Although the static operational constraints in the grid-connected and islanded modes are linear, the investment planning problem described in \eqref{eqn:compact2} is a Mixed-Integer Non-Linear Programming (MINLP) problem due to the inclusion of discrete/continuous investment/operational variables as well as non-linear and non-convex transient security constraints. To solve this intractable MINLP optimization problem, two computationally tractable decomposition strategies based on the notion of Benders decomposition are proposed in this paper, as described in Section \ref{model} in detail.

\section{Two Decomposition Algorithms for Solving the Inertia-Aware MG Planning Problem}\label{model}

\begin{figure}[b!]
    \centering
    \scalebox{0.99}{\includegraphics[width=0.5\textwidth]{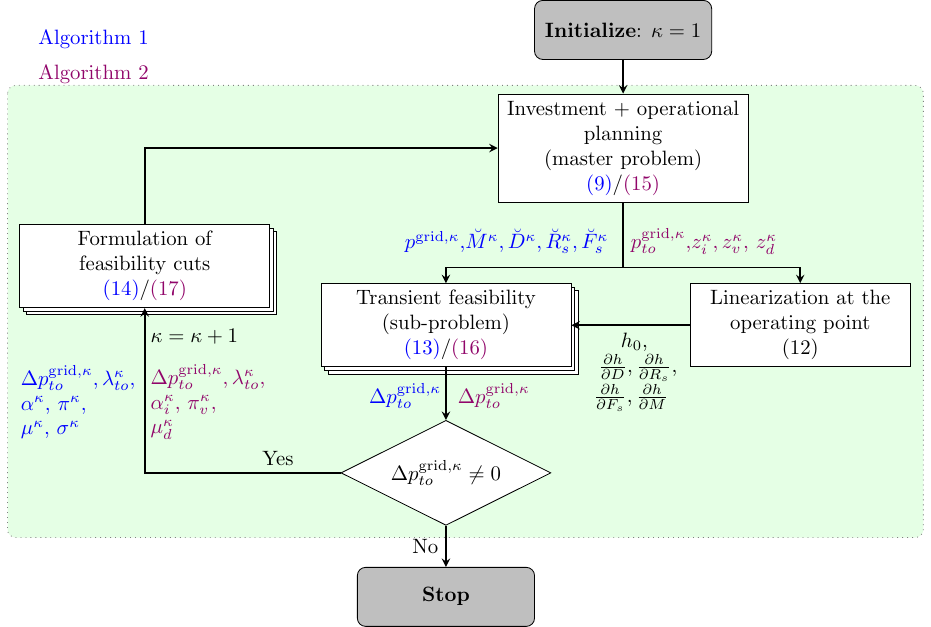}}
    \caption{The proposed decomposition algorithms for the inertia-aware MG  planning problem where variables are differentiated with {A1 (\color{blue}blue}), A2 ({\color{purple}purple}), and both A1 and A2 (black).}
    \label{fig:dec_alg}
\end{figure}
In this section, two decomposition algorithms are proposed, outlining two computationally tractable strategies for solving the inertia-aware MG planning problem in \eqref{eqn:compact2} using dual cutting planes. The solution approach in both algorithms is developed based on the notion of Benders decomposition to tackle the MINLP problem in \eqref{eqn:compact2} through a four-step iterative procedure, as illustrated in Fig.\ref{fig:dec_alg}. In each algorithm, the master problem is related to the MG planning problem under linear static security constraints, while the sub-problem relates to the transient frequency security feasibility. However, the proposed algorithms differ in the formulation of the dual cutting planes as described below, providing different degrees of freedom that can be leveraged in different applications. Unlike the previous work of the authors presented in \cite{9691464}, where the impact of the first-stage investment and operational decisions on the second-stage transient frequency security is not explicitly captured, the dual cutting planes utilized in this paper capture the sensitivity of the solution of the sub-problem to the variables/solution of the master problem. 
These sensitivities are then used in the computation of the master problem at the next iteration, therefore, capturing the impact of the optimal decisions of the master problem on the optimal solution of the sub-problem. Both decomposition algorithms are presented in the sequel. 

\subsection{Algorithm 1} 
In Algorithm 1 (A1), the complicating variables\footnote{Complicating variables are specific variables in a complex optimization problem preventing a tractable distributed solution if appropriate decomposition strategies are not adopted \cite{conejo2006decomposition}.} between the master problem and the sub-problem are the power exchange of the MG with the main grid (i.e., $p^{\mathrm{grid}}_{to}$) and frequency control parameters for different generators (i.e., $\breve M_s$, $\breve D_s$, $\breve F_s $, and $\breve R_s$ for SGs, $\breve M_c$, $\breve D_c$, and $\breve R_c$ for CIGs, and $\breve M$ and $\breve D$ for both SGs and CIGs). The accent $\breve{}$ is used to differentiate a non-normalized parameter $\bullet$ from its normalized counterpart $\breve{\bullet}$. The tasks at each iteration in A1 are detailed as follows:

\subsubsection*{Step 1) Initial Formulation of the Master Problem}
Initially, at iteration $\kappa=1$, the master problem, which is a relaxation of \eqref{eqn:compact2}, is solved to obtain feasible values of the complicating variables. It can be formulated as given below:
\begin{subequations}
\label{eqn:compact-alg1}
\begin{align}
    &\min_{\bm \chi} \;  \Theta^{{\mathrm{inv}}}(\bm \chi^{\mathrm{inv}})+\Theta^{\mathrm{gm,opr}}(\bm \chi^{\mathrm{inv}},\bm \chi^{\mathrm{gm,opr}})+\gamma  \label{eqn:comalg1}\\
    &\textrm{subject to:} \nonumber\\
    &   \gamma\geq  \Theta^{\mathrm{im,opr}}_{to}(\bm \chi^{\mathrm{inv}},\bm \chi^{\mathrm{gm,opr}},\bm \chi^{\mathrm{im,opr}}), \;\;\; \forall t\in\mathcal{T},o\in\mathcal{O},\label{eqn:com alg1}\\
    & \Phi^\mathrm{gm,opr}(\bm \chi^{\mathrm{inv}},\bm \chi^{\mathrm{gm,opr}}) \leq 0, \label{eqn:equal-const alg1}\\
     &\Phi^\mathrm{im,opr}(\bm \chi^{\mathrm{inv}},\bm \chi^{\mathrm{gm,opr}},\bm \chi^{\mathrm{im,opr}}) \leq 0, \label{eqn:inequal-const alg1}\\
     &  {\breve M_s} = \sum_{i\in\mathcal{S}} M_{i}{P_{i}} z_i, \quad  \quad  
     \breve D_s = \sum_{i\in\mathcal{S}} D_{i}P_{i} z_i,\\
    &  \breve R_s = \sum_{i\in\mathcal{S}} \dfrac{K_{i}}{R_{i}}{P_{i}} z_i,\quad\quad\,   \breve F_s = \sum_{i\in\mathcal{S}} \dfrac{K_{i}F_{i}}{R_{i}}{P_{i}} z_i,\\
    & \breve M_c = \sum_{v\in\mathcal{C}^v} M_{v}{P_{c_v}} z_v, \quad   
    \breve D_c = \sum_{v\in\mathcal{C}^v} D_{v}{ P_{c_v}} z_v,\\
    & \breve R_c = \sum_{d \in \mathcal{C}^d}R_{d} {P_{c_d}} z_d,\\
& \breve M = {\breve M_s+ \breve M_c },\\
    &\breve  D = {\breve D_s + \breve D_c + \breve R_c },\\
    &P^{\mathrm{base}} = P_{s}^{\mathrm{base}} + P_{c}^{\mathrm{base}} =  \sum_{i\in\mathcal{S}}  P_{i} z_i + \sum_{c\in\mathcal{C}}  P_{c} z_c.
\end{align}
\end{subequations}
where, the vector $\bm \chi$ includes all investment variables (i.e., $z_i, \forall i \in \mathcal{S}$, $z_v, \forall v \in \mathcal{C}^v$,  $z_d, \forall d \in \mathcal{C}^d$, and $z_c = \{z_v, z_d\}$). The proposed master problem \eqref{eqn:compact-alg1} for A1 as a Mixed-Integer Linear Programming (MILP) problem can be solved by available optimization packages straightforwardly. 

\subsubsection*{Step 2) Linearization at Each Operating Point} 
The nadir constraint \eqref{eq:nadir-const} defined in \eqref{eq:nadir} is highly non-linear and non-convex. To remedy this issue, before its application to the sub-problem, it is linearized around the operating point at each hour of every representative day (i.e., $\forall t \in \mathcal{T}$ and $\forall o \in \mathcal{O}$). For this purpose, $\Delta f_{to}^{\mathrm{max}}$ in \eqref{eq:nadir-const} can be rewritten as follows: 
\begin{align}
&\Delta f^\mathrm{max} =  \dfrac{p^{\mathrm{grid}}_{to}}{h\left(D,R_s, F_s,M\right)},\hspace{3.6cm}
\end{align}
where:
{\small
\begin{align}
{h\left(D,R_s, F_s,M\right)}=\frac{1}{{\frac{1}{D+R_{s}}\left(1+\sqrt{\frac{T(R_{s}-F_{s})}{M}}e^{-\zeta\omega_{n}t_{m}}\right)}}.
\end{align} } 

Now, Taylor expansion is utilized to linearize the nadir constraint \eqref{eq:nadir-const} at each iteration as given below:
\begin{subequations}
    \begin{align}
    & \Tilde{h} \underline{\Delta f}^{\mathrm{max}}\leq p_{to}^{\textrm{grid}}\leq  \Tilde{h} \overline{\Delta f}^{\mathrm{max}},\\
    \begin{split}
    &h \approx \Tilde{h} =  h_{0}^\kappa + \frac{\partial h^{}}{\partial D}(D-D^\kappa) + \frac{\partial  h^{}}{\partial R_s}(R_s-R_{s}^\kappa) \\ 
       &\hspace{1.7cm}  + \frac{\partial  h^{}}{\partial F_s}(F_s-F_{s}^\kappa) + \frac{\partial  h^{}}{\partial M}(M-M^\kappa),
    \end{split}
    \end{align}
\end{subequations}
where, $\Tilde{h}$ is an auxiliary variable, and hereafter, the superscript $\kappa$ is used to differ fixed variables (e.g., $D^\kappa$) from non-fixed variables (e.g., $D$) at iteration $\kappa$ of the decomposition algorithm. The Taylor expansion introduces an approximation error that lowers the accuracy of the expression. The proximity between the true and approximate expressions (i.e., $\Delta f^{\textrm{max},\kappa}_\textrm{exact}$ and $\Delta f^{\textrm{max},\kappa}_\textrm{approx}$, respectively) is computed using the absolute error $\epsilon_\textrm{abs} = | \Delta f_{\textrm{exact}}^{\mathrm{max},\kappa}-\Delta f_{\textrm{approx}}^{\mathrm{max},\kappa} |$ and the relative error $\epsilon_\textrm{rel} = \frac{| \Delta f_{\textrm{exact}}^{\mathrm{max},\kappa}-\Delta f_{\textrm{approx}}^{\mathrm{max},\kappa} |}{\Delta f_{\textrm{exact}}^{\mathrm{max},\kappa}}$. Application of these metrics to the above first-order approximation for 1000 scenarios indicated $\epsilon_\textrm{abs} = 4.7878\times10^{-4}$ and $\epsilon_\textrm{rel} = 0.2734\%$ on average. Although higher order approximations can be adopted for higher accuracy, it can result in the non-linearity and non-convexity of the optimization problem and further complexity. Also, the dynamic simulations indicated in the preceding sections further provide guarantees on the efficacy of the first-order approximation.

\subsubsection*{Step 3) Formulation of the Sub-Problem}
To check whether the transient security constraints \eqref{eq:nadir-const}-\eqref{eq:qss-const} are satisfied based on the solution of the master problem at each iteration or not, feasibility sub-problems are formulated. Given optimal values of complicating variables obtained from solving the master problem at iteration $k$ (e.g., $p^{\mathrm{grid},\kappa}_{to}$), the sub-problem for each hour $t$ of every representative day $o$ can be formulated as a Linear Programming (LP) problem using the slack variable $\Delta p^{\mathrm{grid}}_{to}$ as given below:
\begin{subequations}\label{eq:sub-alg1}
\begin{align}
&\min_{\Delta p_{to}^{\mathrm{grid}}}\;|\Delta p_{to}^{\mathrm{grid}}|\\
&\text{subject to:}\nonumber\\
\begin{split}
    &\Tilde{h} \underline{\Delta f}^{\mathrm{max}} \leq p_{to}^{\mathrm{grid}}+\Delta p_{to}^{\mathrm{grid}}\leq \Tilde{h} \overline{\Delta f}^{\mathrm{max}} , \label{eq:nadir-const2}
\end{split}\\
\begin{split}
    &\Tilde{h} = h_{0}^{\kappa}+\frac{\partial h}{\partial D}\dfrac{(D-\breve{D}^{\kappa})}{P^{\mathrm{base,\kappa}}}+\frac{\partial h}{\partial R_{s}}\dfrac{(R_{s}-\breve{R}_{s}^{\kappa})}{P_{s}^{\mathrm{base,\kappa}}} \\ & \hspace{1.1cm}   +\frac{\partial h}{\partial F_{s}}\dfrac{(F_{s}-\breve{F}_{s}^{\kappa})}{P_{s}^{\mathrm{base,\kappa}}}+\frac{\partial h}{\partial M}\dfrac{(M-\breve{M}^{\kappa})}{P^{\mathrm{base},\kappa}}  , \label{eq:nadir-const22}
\end{split}\\
& \Tilde{i} \underline{\dot{f}}^{\mathrm{max}}\leq p_{to}^{\mathrm{grid}}+\Delta p_{to}^{\mathrm{grid}}\leq \Tilde{i} \overline{\dot{f}}^{\mathrm{max}}, \label{eq:rocof-const2}\\
& \Tilde{i} = \frac{M}{P^{\mathrm{base},\kappa}},\label{eq:rocof-const22}\\
    & \Tilde{j} \underline{\Delta f}^{\mathrm{qss}} \leq p_{to}^{\mathrm{grid}}+\Delta p_{to}^{\mathrm{grid}}\leq \Tilde{j} \overline{\Delta f}^{\mathrm{qss}}, \label{eq:qss-const2}\\
    & \Tilde{j} = \frac{D}{P^{\mathrm{base,\kappa}}}+\frac{{R}_{s}}{P_{s}^{\mathrm{base,\kappa}}}, \label{eq:qss-const22}\\
    & p^{\mathrm{grid}}_{to} = p^{\mathrm{grid},\kappa}_{to} \hspace{3cm} (\text{dual } \lambda_{to}),\\
    &M = {\breve M^{\kappa}}  \hspace{3.7cm}(\text{dual }  \alpha),\\
    &D = {\breve D^{\kappa} }\hspace{3.9cm}(\text{dual }  \pi),\\
    &R_s = {\breve R_{s}}^\kappa \hspace{3.8cm}(\text{dual }  \mu),\\
    &F_s = {\breve F_{s}}^\kappa \hspace{3.9cm}(\text{dual }  \sigma),
\end{align}
\end{subequations}
where, the auxiliary variables $\Tilde{h}$ in \eqref{eq:nadir-const2}-\eqref{eq:nadir-const22}, $\Tilde{i}$ in \eqref{eq:rocof-const2}-\eqref{eq:rocof-const22}, and $\Tilde{j}$ in \eqref{eq:qss-const2}-\eqref{eq:qss-const22} are used to include the constraints \eqref{eq:nadir-const}, \eqref{eq:rocof-const}, and \eqref{eq:qss-const}, respectively, in the sub-problem without any non-linear term. Also, $\lambda_{to}$, $\alpha$, $\pi$, $\mu$ and $\sigma$ are dual variables for the constraints fixing the complicating variables (i.e., the power exchange with the main grid, aggregated inertia, damping, droop, and turbine power fraction, respectively) in the sub-problem. These dual variables provide the sensitivity of the optimal solution obtained from solving the sub-problem to the optimal values of the complicating variables obtained from solving the master problem. 

If the optimal solution in \eqref{eq:sub-alg1} is equal to zero for all hours of all representative days, this implies the feasibility of the master problem. In this case, A1 is terminated, and the solution of the master problem at the final iteration is the optimal solution. On the other hand, if the optimal solution in \eqref{eq:sub-alg1} is non-zero for even one hour of a specific representative day, i.e., $|\Delta p^{\mathrm{grid}}_{to}|> 0$, this implies the infeasibility of the sub-problem given the values of the complicating variables. Physically, this is associated with violations of transient security constraints. To eliminate these violations, feasibility cuts are added to the master problem. 

\subsubsection*{Step 4: Formulation of Resilient Feasibility Cut}
Given the dual variables obtained from solving each sub-problem at iteration $\kappa$ (i.e., $\lambda_{to}^\kappa$, $\alpha^\kappa$, $\pi^\kappa$, $\mu^\kappa$, and $\sigma^\kappa$), the master problem in \eqref{eqn:compact-alg1} is updated with the dual cutting planes if the sub-problem in \eqref{eq:sub-alg1} is infeasible, i.e., $|\Delta p^{\mathrm{grid},\kappa}_{to}| >0$. The cutting planes added to the master problem are defined as follows:
\begin{align}
\begin{split}
    &\Delta p_{to}^{\mathrm{grid,\kappa}}+\lambda_{to}^{\kappa}(p_{to}^{\mathrm{grid}}-p^{\mathrm{grid},\kappa}_{to})+\alpha^{\kappa}(\breve{M}-\breve{M}^\kappa) \\
    &\hspace{1.2cm}+\pi^{\kappa}(\breve{D}-\breve{D}^{\kappa})+\mu^{\kappa}(\breve{R}_{s}-\breve{R}_{s}^{\kappa}) \\
    &\hspace{1.2cm}+\sigma^{\kappa}(\breve{F}_{s}-\breve{F}_{s}^{\kappa}) \leq 0, \hspace{1cm}\forall t,\forall o,\forall \kappa
    \end{split}
\end{align}

The dual cutting planes in A1 are associated with the grid power exchange and the unit control parameters. This implies that the sufficiency of frequency support is examined based on the level of the power exchange with the main grid ($p^{\mathrm{grid}}_{to}$) in addition to the aggregated levels of inertia ($M$) and damping ($D$) of CIG and SG units plus the droop support ($R_s$) and the turbine power fraction ($F_s$) of SG units.

\subsection{Algorithm 2}
Differing from A1, the complicating variables in Algorithm 2 (A2) are defined by the investment status of different units (i.e., $z_i, \forall i \in \mathcal{S}$, $z_v, \forall v \in \mathcal{C}^v$,  $z_d, \forall d \in \mathcal{C}^d$, and $z_c = \{z_v, z_d\}$) and the power exchange with the main grid (i.e., $p^{\mathrm{grid},\kappa}_{to}, \forall t \in \mathcal{T},\forall o \in \mathcal{O}$). 
The tasks at each iteration in A2 are detailed as follows:

\subsubsection*{Step 1: Initial Formulation of the Master Problem}
 At iteration $\kappa=1$, the feasible values of the complicating variables are obtained from the MILP master problem as a relaxation of \eqref{eqn:compact2} formulated as follows:
\begin{subequations}
\label{eqn:compact-alg2}
\begin{align}
&\hspace{-1.5cm} \min_{\bm \chi} \;   \Theta^{{\mathrm{inv}}}(\bm \chi^{\mathrm{inv}})+\Theta^{\mathrm{gm,opr}}(\bm \chi^{\mathrm{inv}},\bm \chi^{\mathrm{gm,opr}})+\gamma \label{eqn:comalg2}\\
\;\textrm{subject to:}\nonumber\\ 
&\hspace{-1.5cm} \gamma\geq  \Theta^{\mathrm{im,opr}}_{to}(\bm \chi^{\mathrm{inv}},\bm \chi^{\mathrm{gm,opr}},\bm \chi^{\mathrm{im,opr}}), \; \forall t\in\mathcal{T},o\in\mathcal{O},\label{eqn:com alg2}\\
&\hspace{-1.5cm} \Phi^\mathrm{gm,opr}(\bm \chi^{\mathrm{inv}},\bm \chi^{\mathrm{gm,opr}}) \leq 0, \label{eqn:equal-const alg2}\\
&\hspace{-1.5cm} \Phi^\mathrm{im,opr}(\bm \chi^{\mathrm{inv}},\bm \chi^{\mathrm{gm,opr}},\bm \chi^{\mathrm{im,opr}}) \leq 0, \label{eqn:inequal-const alg2}\\
&\hspace{-1.5cm} P^{\mathrm{base}} = P_{s}^{\mathrm{base}} + P_{c}^{\mathrm{base}} =   \sum_{i\in\mathcal{S}}  P_{i} z_i + \sum_{c\in\mathcal{C}}  P_{c} z_c.
\end{align}
\end{subequations}

\subsubsection*{Step 2: Linearization}
The linearization step is undertaken in a similar manner to A1 and the result is applied to the sub-problem.

\subsubsection*{Step 3: Formulation of the Sub-Problem}
The feasibility of the master problem defined in \eqref{eqn:compact-alg2} is evaluated using the sub-problems described as follows:
\begin{subequations}\label{eq:sub-alg2}
\begin{align}
&\min_{\Delta  p^{\mathrm{grid},\kappa}_{to}} \; |\Delta  p^{\mathrm{grid},\kappa}_{to}|\\
&\text{s.t.}\nonumber\\
\begin{split}
    &\Tilde{h} \underline{\Delta f}^{\mathrm{max}} \leq p_{to}^{\mathrm{grid}}+\Delta p_{to}^{\mathrm{grid}}\leq \Tilde{h} \overline{\Delta f}^{\mathrm{max}} , 
\end{split}\\
\begin{split}
    &\Tilde{h}=h_{0}^{\kappa}+\frac{\partial h}{\partial D}(D-\breve{D}^{\kappa})+\frac{\partial h}{\partial R_{s}}(R_{s}-\breve{R}_{s}^{\kappa})\\
    &\hspace{1.1cm}+\frac{\partial h}{\partial F_{s}}(F_{s}-\breve{F}_{s}^{\kappa})+\frac{\partial h}{\partial M}(M-\breve{M}^{\kappa})  , 
\end{split}\\
& \Tilde{i} \underline{\dot{f}}^{\mathrm{max}}\leq p_{to}^{\mathrm{grid}}+\Delta p_{to}^{\mathrm{grid}}\leq \Tilde{i} \overline{\dot{f}}^{\mathrm{max}},\\
& \Tilde{i} = M,\\
& \Tilde{j} \underline{\Delta f}^{\mathrm{qss}} \leq p_{to}^{\mathrm{grid}}+\Delta p_{to}^{\mathrm{grid}}\leq \Tilde{j} \overline{\Delta f}^{\mathrm{qss}}, \\
& \Tilde{j} = D+{R}_{s},\\
&M_{s}=\sum_{i\in\mathcal{S}}M_{i}\dfrac{P_{i}}{P_{s}^{\mathrm{base,\kappa}}}z_{i},\;\;\;\, D_{s}=\sum_{i\in\mathcal{S}}D_{i}\dfrac{P_{i}}{P_{s}^{\mathrm{base,\kappa}}}z_{i},\\    
&R_{s}=\sum_{i\in\mathcal{S}}\dfrac{K_{i}}{R_{i}}\dfrac{P_{i}}{P_{s}^{\mathrm{base,\kappa}}}z_{i},\quad\, F_{s}=\sum_{i\in\mathcal{S}}\dfrac{K_{i}F_{i}}{R_{i}}\dfrac{P_{i}}{P_{s}^{\mathrm{base,\kappa}}}z_{i},\\
&M_{c}=\sum_{v\in\mathcal{C}^{v}}M_{v}\dfrac{P_{c_{v}}}{P_{c}^{\mathrm{base,\kappa}}}z_{v},\; D_{c}=\sum_{v\in\mathcal{C}^{v}}D_{v}\dfrac{P_{c_{v}}}{P_{c}^{\mathrm{base,\kappa}}}z_{v},\\
&R_{c}=\sum_{d\in\mathcal{C}^{d}}R_{d}\dfrac{P_{c_{d}}}{P_{c}^{\mathrm{base,\kappa}}}z_{d},\\
&M=\dfrac{M_{s}P_{s}^{\mathrm{base,\kappa}}+M_{c}P_{c}^{\mathrm{base,\kappa}}}{P_{s}^{\mathrm{base,\kappa}}+P_{c}^{\mathrm{base,\kappa}}},\\
&D=\dfrac{D_{s}P_{s}^{\mathrm{base,\kappa}}+D_{c}P_{c}^{\mathrm{base,\kappa}}+R_{c}P_{c}^{\mathrm{base,\kappa}}}{P_{s}^{\mathrm{base,\kappa}}+P_{c}^{\mathrm{base,\kappa}}},\\
& p^{\mathrm{grid}}_{to} = p^{\mathrm{grid},\kappa}_{to} \hspace{3.5cm} (\text{dual } \lambda_{to}),\\
&z_i = z_{i}^\kappa \hspace{1.7cm}\forall i \in \mathcal{S} \hspace{1.8cm}(\text{dual } \alpha_{i}),\\
&z_v = z_{v}^\kappa\hspace{1.7cm}\forall v \in \mathcal{C}^v \hspace{1.6cm}(\text{dual } \pi_{v}),\\
&z_d = z_{d}^\kappa\hspace{1.7cm}\forall d \in \mathcal{C}^d \hspace{1.6cm}(\text{dual } \mu_{d}),
\end{align}
\end{subequations}
where, $\lambda_{to}$, $\alpha_{i}$, $\pi_{v}$, and $\mu_{d}$ are dual variables for the constraints fixing the complicating variables (i.e., the power exchange with the main grid, and investment status of SG and CIG units with VSM and droop control, respectively) in the sub-problem.

\subsubsection*{Step 4: Formulation of Resilient Feasibility Cut}
Given the dual variables obtained from solving each sub-problem at iteration $\kappa$ (i.e., $\lambda_{to}^\kappa$, $\alpha^\kappa_i$, $\pi^\kappa_v$, and $\mu^\kappa_d$), the master problem in \eqref{eqn:compact-alg2} is updated with the dual cutting planes if the sub-problem in \eqref{eq:sub-alg2} is infeasible, i.e., $|\Delta p^{\mathrm{grid},\kappa}_{to}| >0$. The cutting planes added to the master problem are defined as follows:
\begin{align}
\begin{split}
    \Delta p_{to}^{\mathrm{grid,\kappa}}&+\lambda_{to}^{\kappa}(p_{to}^{\mathrm{grid}}-p_{to}^{\mathrm{grid,\kappa}})+\sum_{i\in\mathcal{S}}\alpha_{i}^{\kappa}(z_{i}-z_{i}^{\kappa})  \\&  +\sum_{v\in\mathcal{C}^{v}}\pi_{v}^{\kappa}(z_{v}-z_{v}^{\kappa})+\sum_{d\in\mathcal{C}^{d}}\mu_{d}^{\kappa}(z_{d}-z_{d}^{\kappa})\leq0,\hspace{0.1cm}\forall t,\forall o.
    \end{split}
\end{align}

The dual cutting planes in A2 are associated with the grid power exchange and the investment status of different units. Therefore, the feasibility and sufficiency of the frequency support depend on the level of the power exchange with the main grid ($p^{\mathrm{grid}}_{to}$) and the level of support offered by different invested units ($z_i$, $z_v$, and $z_d$).

Both models are able to obtain an optimal solution as discussed in Section \ref{study}.
In both A1 and A2, the master problem is MILP while the sub-problem is LP. These are tractable reformulations of the MINLP problem in \eqref{eqn:compact2} and can be easily solved with available off-the-shelf optimization solvers. Fig.~\ref{fig:dec_alg} summarizes the proposed algorithms.

\section{Case Study Results}\label{study}
In this section, the proposed algorithms are tested and verified on an 18-bus LV network and a 30-bus MV network under various operating scenarios. The implementation is done in \textsc{MATLAB}, with the optimization model formulated in \textsc{YALMIP}~\cite{Lofberg2004} and solved by \textsc{Gurobi}~\cite{gurobi}.

It is assumed that the 18-bus network has one SG unit already present and the investment candidates comprise of one SG unit ($\mathrm{SG}_2$), two {\it grid-supporting} Photovoltaic (PV) CIG units ($\mathrm{PV}_1$, $\mathrm{PV}_2$), and one PV CIG unit ($\mathrm{PV}_3$) operating in {\it grid-feeding} mode with fixed power output. Candidates $\mathrm{PV}_1$ and $\mathrm{PV}_2$ provide VSM and droop control, respectively. The generator parameters are as described in Table \ref{tab:gen_param}. The transient frequency security constraints are enforced through thresholds imposed on frequency nadir $(\overline{{\Delta f}}^\textrm{max} = - \underline{{\Delta f}}^\textrm{max} = 0.6\,\mathrm{Hz})$, RoCoF $(\overline{\dot{f}}^\textrm{max}= - \underline{\dot{f}}^\textrm{max} = 2\,\mathrm{Hz/s})$, and quasi-steady-state frequency deviation $(\overline{\Delta f}^\textrm{qss} = - \underline{\Delta f}^\textrm{qss} = 0.2\,\mathrm{Hz})$. Further network parameters can be found in \cite{9691464}. In the following, the performance of the proposed algorithms is compared with the three-stage algorithm introduced by the authors previously in \cite{9691464}, herein referred to as Algorithm 0 (A0). Additionally, the \textit{Base} case study is used to denote the model without transient frequency security constraints.

\newcolumntype{g}{>{\columncolor{green!10}}c}
\newcolumntype{x}{>{\columncolor{orange!20}}c}
\begin{table}[!t]
    \renewcommand{\arraystretch}{1.2}
    \centering
    \caption{Generator Control Parameters and Investment Costs}
    \label{tab:gen_param}
    \resizebox{0.95\linewidth}{!}{%
     \begin{tabular}{c|g|x|x|x|x}
      \toprule
         &  $\mathrm{SG}_1$ &  $\mathrm{SG}_2$ &  $\mathrm{PV}_1$ &  $\mathrm{PV}_2$ &  $\mathrm{PV}_3$   \\
    \midrule\midrule
        {Annualized} investment cost $(\$)$ &  - &  $40\,000$  &  $70\,000$ &  $65\,000$ &  $60\,000$\\
            Capacity $(\SI{}{\kilo\watt})$  &  $280$  &  $350$ &  $350$ &  $350$ &  $350$ \\
    $M\,(\SI{}{\second})$ &  $14$ &  $14$ &  $14$  &  - &  -\\
    $D\,(\mathrm{p.u.})$ &  $0.9$ &  $0.9$ &  $0.9$ &  - &  -\\
    $K\,(\mathrm{p.u.})$ &  $1$ &  $1$ &  $1$  &  $1$ &  -\\
    $R\,(\mathrm{p.u.})$ &  $0.03$ &  $0.03$ &  - &  $0.05$ &  -\\
    $F\,(\mathrm{p.u.})$ &  $0.35$ &  $0.35$ &  - &  - &  -\\
    \bottomrule 
    \rowcolor{green!10}
    \multicolumn{6}{@{}p{1.5in}}{\footnotesize Existing generator}\\
    \rowcolor{orange!20}
    \multicolumn{6}{@{}p{1.5in}}{\footnotesize Candidate generators}
    \end{tabular}
 }
   \vspace{-1.2em}
   \end{table}


\subsection{Planning Costs}\label{sec:planning-costs}

\begin{table}[!b]
 \renewcommand{\arraystretch}{1.2}
 \vspace{-0.0em}
    \centering
    \caption{Comparison of optimal costs and decisions, inertia support, and computational performance for each algorithm using four representative days.}
    \label{tab:cost-init}
    \resizebox{1\linewidth}{!}{
\begin{tabular}{l|cccc|}
\cline{2-5}
 & \multicolumn{1}{c|}{\textbf{Base}} & \multicolumn{1}{c|}{\textbf{Algorithm 0}} & \multicolumn{1}{c|}{\textbf{Algorithm 1}} & \textbf{Algorithm 2 }\\ \toprule\hline
 \multicolumn{5}{|c|}{Costs and decisions} \\ \hline
\multicolumn{1}{|l|}{Total cost (\$)} & \multicolumn{1}{c|}{223390} & \multicolumn{1}{c|}{244780} & \multicolumn{1}{c|}{242740} & 242740 \\ \hline
\multicolumn{1}{|l|}{Investment cost (\$)} & \multicolumn{1}{c|}{61000} & \multicolumn{1}{c|}{126000} & \multicolumn{1}{c|}{131000} & 131000 \\ \hline
\multicolumn{1}{|l|}{Investment decisions} & \multicolumn{1}{c|}{PV$_3$}& \multicolumn{1}{c|}{PV$_2$, PV$_3$ }& \multicolumn{1}{c|}{PV$_1$, PV$_3$} & \multicolumn{1}{c|}{PV$_1$, PV$_3$} \\ \hline
\multicolumn{1}{|l|}{Operational cost (\$)} & \multicolumn{1}{c|}{162390} & \multicolumn{1}{c|}{118780} & \multicolumn{1}{c|}{111740} & 111740 \\ \hline
\multicolumn{1}{|l|}{\begin{tabular}[c]{@{}c@{}}Demand shift\\ \addlinespace[-3.5pt] penalty  (\$)\end{tabular}} & \multicolumn{1}{c|}{3675} & \multicolumn{1}{c|}{8468} & \multicolumn{1}{c|}{7787} & 7787 \\ \hline
\multicolumn{1}{|l|}{\begin{tabular}[c]{@{}c@{}}Demand disconnection\\ \addlinespace[-3.5pt] penalty  (\$)\end{tabular}} & \multicolumn{1}{c|}{14536} & \multicolumn{1}{c|}{5337} & \multicolumn{1}{c|}{5337} & 5337 \\ \hline\bottomrule
\multicolumn{5}{|c|}{Frequency support} \\ \hline
\multicolumn{1}{|l|}{M (s)} & \multicolumn{1}{c|}{7.84} & \multicolumn{1}{c|}{7.84} & \multicolumn{1}{c|}{17.64} & 17.64 \\ \hline
\multicolumn{1}{|l|}{D (p.u)} & \multicolumn{1}{c|}{0.50} & \multicolumn{1}{c|}{18.00} & \multicolumn{1}{c|}{1.13} & 1.13 \\ \hline  \bottomrule
\multicolumn{5}{|c|}{Computational performance} \\ \hline
\multicolumn{1}{|l|}{Number of iterations} & \multicolumn{1}{c|}{-} & \multicolumn{1}{c|}{6} & \multicolumn{1}{c|}{4} & 4 \\ \hline
\multicolumn{1}{|l|}{Computation   time (s)} & \multicolumn{1}{c|}{612} & \multicolumn{1}{c|}{4540} & \multicolumn{1}{c|}{3438} & 3386 \\ \hline 
\end{tabular}}
\end{table}

The costs and planning decisions considering four representative days are compared with each of the three algorithms A0, A1, and A2 as indicated in Table \ref{tab:cost-init}. With all algorithms, there exists an increment in total costs compared to the Base case. For A0, a 10\% increment in total costs is observed, compared to an 8.8\% increment for A1 and A2. The total investment costs are lowest with A0 compared to A1 and A2 while the solution for A0 results in the highest operational costs.

When transient frequency security constraints are applied to the problem (A0, A1, and A2), it is essential that the algorithm minimizes costs while ensuring that the level of frequency support in the network is adequate to eliminate violations. Recall that the transient frequency response depends on the aggregated levels of parameters $M,\,D,\,R_s$, and $F_s$ provided by the installed units (see \eqref{eqn: Msup} and \eqref{eqn: Dsup}). While the installed capacity is similar for algorithms A0, A1, and A2, as depicted in Table \ref{tab:cost-init}, the support offered by the selected CIG units varies.

Compared to the Base case, A0 selects an additional droop-based CIG unit ($\mathrm{PV}_2$) only contributing to the aggregated damping level while A1 and A2 select an additional VSM-based CIG unit ($\mathrm{PV}_1$) contributing to both the aggregated damping and inertia levels (see \eqref{eqn: Msup} and \eqref{eqn: Dsup}). Droop-based CIG units contribute to frequency support only in the region of primary frequency response and not during inertia response. As more frequency support is available from the units selected by A1 and A2, the preventive operational actions are kept to a minimum. The solution provided by A0 considers the cheapest investment unit while A1 and A2 select the most \textit{cost-effective} unit. Unlike A0, the decomposition approach used in A1 and A2 provides sensitivity information from the sub-problem to the master problem resulting in a solution that is not only optimal in cost but also ensures optimal frequency support.

\subsection{Dynamic Performance}

\begin{figure}[t!]
\centering
    \begin{subfigure}[t]{1\linewidth}
        \centering
        \includegraphics[width=.70\textwidth]{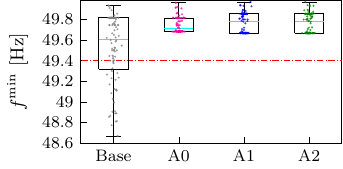}\\
        \caption{}
        \label{fig:met_comp b}
    \end{subfigure}%
   \\
     \begin{subfigure}[t]{1\linewidth}
        \centering
        \includegraphics[width=0.66\textwidth]{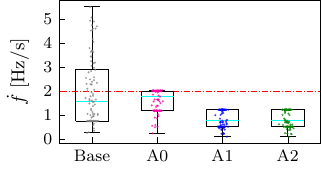}
        \caption{}
        \label{fig:met_comp a}
    \end{subfigure}%
\\
    \begin{subfigure}[t]{1\linewidth}
        \centering
        \includegraphics[width=.70\textwidth]{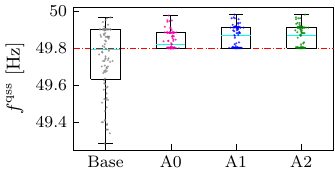}\\
        \caption{}
        \label{fig:met_comp c}
    \end{subfigure}
    \caption{Metric variation in each algorithm with respect to the (a) nadir, (b) RoCoF, and (c) quasi-steady-state frequency considering all hours in four representative days.}
    \label{fig:met_comp}
    \vspace{-1.2em}
\end{figure}

Based on the units installed by each algorithm, the total aggregated level of $M$ and $D$ are 7.84~s and 18~p.u. for A0 as compared to 17.64~s and 1.13~p.u. for A1 and A2. Fig. \ref{fig:met_comp} presents a box plot that indicates the variations in the measured values for each of the frequency security metrics for the 96 hours in four representative days. The security limit in each case is indicated by the dotted red line. 

In the case of the nadir values, averages of 49.77 Hz and 49.70 Hz are recorded in A0 and A1/A2, respectively (see Fig.~\ref{fig:met_comp b}), as compared to the Base case where an average of 49.55 is obtained. The variation of nadir in A0  indicates a smaller range as compared to A1/A2 with a slightly improved performance. In the case of the RoCoF values, depicted in Fig.~\ref{fig:met_comp a}, an average of 1.59~Hz/s is achieved for A0 as compared to 0.79 Hz/s for A1 and A2. RoCoF is mainly dependent on the total inertia level ($M$) present in the network (see \eqref{eq:rocof}). The solution provided by A0 provides inertia levels of 7.84~s, provided mainly by the pre-installed SG. The results based on A1 and A2 include the additional installation of VSM-based unit $\mathrm{PV}_1$ resulting in an inertia level of 17.64~s, and therefore, a better performance level compared to A0. On the other hand, the quasi-steady-state frequency is dependent on the aggregated $D$ and $R_s$ parameters, as indicated in \eqref{eq:qss}. In Fig.~\ref{fig:met_comp c}, averages of 49.89~Hz and 49.87~Hz are observed for A0 and A1/A2, respectively. From Table~\ref{tab:cost-init}, it can be seen that the aggregated damping levels are higher with the units installed by the A0 solution hence providing slightly better performance.

\begin{figure}[t!]
    \centering
    \includegraphics[width=0.45\textwidth]{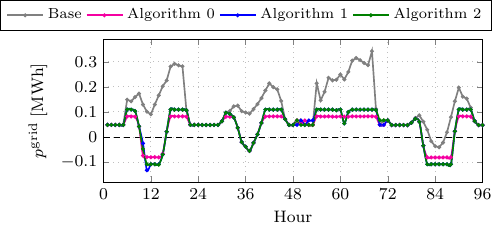}
    \caption{Impact of the transient frequency constraints on active power exchange with the main grid for the different algorithms (-/+ indicate power export/import).}
    \label{fig:p_exc}
\end{figure}

The impact of the inclusion of transient constraints on the active power exchange with the main grid is shown in Fig.~\ref{fig:p_exc}. All the algorithms provide a solution that is robust to the loss of power exchange with the main grid in every operation scenario. For the MG, the power exchange with the main grid is usually the largest power injection, the loss of this power exchange may result in large frequency excursions. A0 explicitly restricts the bounds on the grid power exchange at each hour while A1 and A2 vary the dispatch based on the sensitivities to both the power loss and available inertia. The result of A0 is shown to be more conservative compared to A1 and A2. This is especially due to the lack of bidirectional information exchange between the problems at each stage. A0 only restricts power exchange based on the feasibility of the transient frequency security problem, without knowledge of the impact of different control parameters on the transient frequency security. It is noteworthy to mention that during instances of power import, the disconnection from the main grid can result in potential under-frequency, while for power export, over-frequencies can be recorded.

\begin{figure}[t!]
    \centering
    \includegraphics[width=0.45\textwidth]{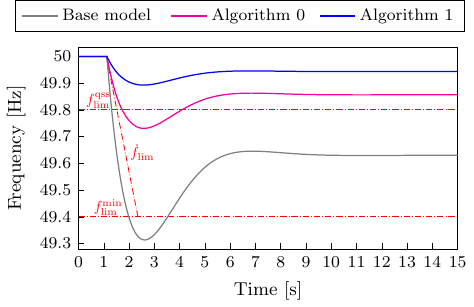}
    \caption{Evolution of the CoI frequency for 15 seconds after disconnection of the MG from the main grid at hour 68.}
    \label{fig:dyn_sim}
    \vspace{-1.5em}
\end{figure}

The analytical performance, shown in Fig.~\ref{fig:met_comp}, is further validated using time-domain simulations. Given the optimal solution provided by the Base case, A0, and A1 (A2 is committed since it has the same solution as A1), Fig. \ref{fig:dyn_sim} indicates the frequency trajectories for the operational scenario at hour 68 out of 96 operational hours in four representative days. Note that based on Fig.~\ref{fig:p_exc}, this hour represents the highest power exchange from the grid, and thus, the worst-case mismatch in power if the MG is disconnected. The superiority of A1 over A0 is further validated in Fig.~\ref{fig:dyn_sim} which compares the frequency trajectories of each technique using a time domain simulation. The dynamic simulation is performed with PyRAMSES~\cite{7365497} with the grid disconnection occurring at time $t= 1$~s and the dotted red lines have been used to indicate the transient frequency security limits. 
 While in this scenario, all metrics of A1 show a better performance than A0, this may however not always be the case as indicated in Fig.~\ref{fig:met_comp}. The frequency evolution is dependent on both the level of power mismatch at the instant of disconnection dictated by the hourly power exchange and the control support available. 

\subsection{Computational Performance}
The computation time for each of the methods is indicated in Table \ref{tab:cost-init}. The solution with A0 is obtained after six iterations as compared to only four iterations required for A1 and A2. The dual cutting planes from all complicating variables ensure faster elimination of the infeasible regions in A1 and A2 as opposed to only imposing limits on the level of power exchange with the main grid in A0. Therefore, the convergence of A1 and A2 is shown to be much faster than A0. Furthermore, a 25\% decrease in computational time is recorded with the application of A1 and A2. While A1 and A2 provide similar optimal solutions, A2 is shown to obtain the solution faster as compared to A1 and is therefore more computationally efficient.

\subsection{Sensitivity to Variation in Security Limits}

Tightening the limits, i.e., reducing the upper bounds and increasing the lower bounds, increases the system requirement on power reserves necessary for frequency support from the system. 
In this paper, three case studies are defined to analyze the effect of threshold variations, i.e.:
\begin{enumerate}
    \item Case A: denotes the initially considered limits
    \item Case B: denotes the tightening of \textit{only} the RoCoF limit
    \item Case C: denotes the tightening of \textit{only} the quasi-steady-state frequency limit
\end{enumerate}
\begin{figure}[t!]
    \centering
    \includegraphics[width=0.35\textwidth]{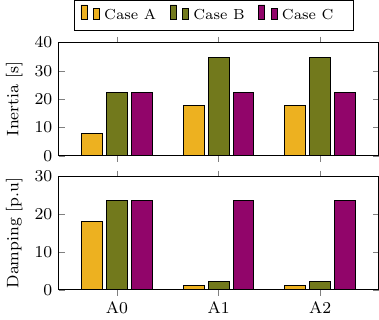}
    \caption{Variation of the normalized aggregated inertia and damping constants for different limits on the frequency security metrics.}
    \label{fig:freq_supcomp}
\end{figure}

 Higher support requirements can be met by further leveraging high-cost preventive actions. If, however, the current system configuration fails to meet the security requirements, new units can be installed to increase support levels. Furthermore, as shown in \eqref{eq:nadir}-\eqref{eq:qss}, the different metrics are dependent on either one or a combination of different control parameters. Hence, the commitment of a generator will depend on its suitability to enhance performance. The effect of the limit variation on the aggregated damping and inertia magnitudes is presented in Fig.~\ref{fig:freq_supcomp}, while the sensitivity of the planning solutions to threshold variations is presented in Tables~\ref{tab:roco18-1}~and~\ref{tab:qss_18_1}.
  \begin{table}[!t]
        \centering
        \caption{Planning costs and decisions with tighter transient security bounds on RoCoF (Case B: Tightening RoCoF to 0.5 Hz/s).}
        \renewcommand{\arraystretch}{1.2}
        \resizebox{1\linewidth}{!}{
        \begin{tabular}{lccc|}
        \cline{2-4}
        \multicolumn{1}{l|}{} & \multicolumn{1}{c|}{\textbf{Algorithm 0}} & \multicolumn{1}{c|}{\textbf{Algorithm 1}} & \textbf{Algorithm 2} \\ \toprule\hline
        \multicolumn{4}{|c|}{Costs and decisions} \\ \hline
        \multicolumn{1}{|l|}{Total cost (\$)} & \multicolumn{1}{c|}{302790} & \multicolumn{1}{c|}{295303} & 295319 \\ \hline
        \multicolumn{1}{|l|}{Investment cost (\$)} & \multicolumn{1}{c|}{166000} & \multicolumn{1}{c|}{170000} & 170000 \\ \hline
        \multicolumn{1}{|l|}{Investment decisions} & \multicolumn{1}{c|}{\begin{tabular}[c]{@{}c@{}}PV$_2$, PV$_3$,\\  SG$_2$\end{tabular}} & \multicolumn{1}{c|}{\begin{tabular}[c]{@{}c@{}}PV$_1$, PV$_3$,\\ SG$_2$\end{tabular}} & \begin{tabular}[c]{@{}c@{}}PV$_1$, PV$_3$,\\ SG$_2$\end{tabular} \\ \hline
        \multicolumn{1}{|l|}{Operational cost (\$)} & \multicolumn{1}{c|}{136790} & \multicolumn{1}{c|}{125303} & 125319 \\ \hline
        \multicolumn{1}{|l|}{Demand shift penalty (\$)} & \multicolumn{1}{c|}{15765} & \multicolumn{1}{c|}{6437} & 6437 \\ \hline
        \multicolumn{1}{|l|}{Demand disconnection penalty (\$)} & \multicolumn{1}{c|}{0} & \multicolumn{1}{c|}{0} & 0 \\ \hline \bottomrule
        \multicolumn{4}{|c|}{Computational performance} \\ \hline
        \multicolumn{1}{|l|}{No. of iterations} & \multicolumn{1}{c|}{6} & \multicolumn{1}{c|}{4} & 4 \\ \hline
        \multicolumn{1}{|l|}{Computation time (s)} & \multicolumn{1}{c|}{3484} & \multicolumn{1}{c|}{2453} & 2300 \\ \hline 
        \end{tabular}}
      \label{tab:roco18-1}
    \end{table}
    
 \begin{table}[!t]
        \centering
        \caption{Planning costs and decisions with tighter transient security bounds on quasi-steady-state frequency (Case C: Tightening quasi-steady-state frequency deviation limit to 0.1 Hz). } 
        \renewcommand{\arraystretch}{1.2}
        \resizebox{1\linewidth}{!}{
        \begin{tabular}{lccc|}
        \cline{2-4}
        \multicolumn{1}{l|}{} & \multicolumn{1}{c|}{\textbf{Algorithm 0}} & \multicolumn{1}{c|}{\textbf{Algorithm 1}} & \textbf{Algorithm 2} \\ \hline
        \multicolumn{4}{|c|}{Costs and decisions} \\ \toprule\hline
        \multicolumn{1}{|l|}{Total cost (\$)} & \multicolumn{1}{c|}{296734} & \multicolumn{1}{c|}{296665} & 296665 \\ \hline
        \multicolumn{1}{|l|}{Investment cost (\$)} & \multicolumn{1}{c|}{166000} & \multicolumn{1}{c|}{166000} & 166000 \\ \hline
        \multicolumn{1}{|l|}{Investment decisions} & \multicolumn{1}{c|}{\begin{tabular}[c]{@{}c@{}}PV$_2$, PV$_3$, \\ SG$_2$\end{tabular}} & \multicolumn{1}{c|}{\begin{tabular}[c]{@{}c@{}}PV$_2$, PV$_3$,\\  SG$_2$\end{tabular}} & \begin{tabular}[c]{@{}c@{}}PV$_2$, PV$_3$,\\  SG$_2$\end{tabular} \\ \hline
        \multicolumn{1}{|l|}{Operational cost (\$)} & \multicolumn{1}{c|}{130734} & \multicolumn{1}{c|}{130665} & 130665 \\ \hline
        \multicolumn{1}{|l|}{Demand shift penalty (\$)} & \multicolumn{1}{c|}{14786} & \multicolumn{1}{c|}{14756} & 14756 \\ \hline
        \multicolumn{1}{|l|}{Demand disconnection penalty (\$)} & \multicolumn{1}{c|}{0} & \multicolumn{1}{c|}{0} & 0 \\ \hline \bottomrule
        \multicolumn{4}{|c|}{Computational   performance} \\ \hline
        \multicolumn{1}{|l|}{No. of iterations} & \multicolumn{1}{c|}{7} & \multicolumn{1}{c|}{3} & 3 \\ \hline
        \multicolumn{1}{|l|}{Computation time} & \multicolumn{1}{c|}{3787} & \multicolumn{1}{c|}{1956} & 1947 \\ \hline 
        \end{tabular}}
       \label{tab:qss_18_1}
       \vspace{-1.2em}
     \end{table}

In Case B, the RoCoF thresholds are reduced from 2 Hz/s to 0.5 Hz/s. From \eqref{eq:rocof}, RoCoF is more dependent on the aggregated inertia $M$. Fig. \ref{fig:freq_supcomp} indicates an increase in inertia levels for all algorithms. A significant increment to 34.66 s with A1/A2 as compared to 17.64 s in Case A is shown while a lower increment of 22.28 s is obtained with A0. While all algorithms present the need to install an additional unit as indicated in Table \ref{tab:roco18-1}, units installed with A0 include a droop-based generator that has no contribution to the aggregated inertia $M$. Hence, A0 resorts to more expensive preventative actions to further eliminate transient security violations increasing total costs. 

In Case C, the quasi-steady-state frequency deviation bound is reduced from 0.2 Hz to 0.1 Hz. From \eqref{eq:qss}, the quasi-steady-state frequency is dependent on the magnitude of control parameters for damping and droop. In Fig.~\ref{fig:freq_supcomp}, damping levels for Case C with A1/A2 are shown to increase from 1.13 p.u. in Case A to 23.54 p.u. in Case C. Note that in Case C, the units installed with A1/A2 include a droop-based CIG PV$_2$ instead of VSM-based PV$_1$ (see Table \ref{tab:qss_18_1}) as in Case A and B. This is due to the higher requirement for damping support with Case C as compared to Case A necessitating the adoption of units that result in better performance. 

The results presented in Tables \ref{tab:roco18-1} and \ref{tab:qss_18_1} as well as Fig. \ref{fig:freq_supcomp} further indicate the need for sensitivity information exchange between the two stages of the algorithm, i.e., the master problem and the sub-problem. This simultaneously optimizes both frequency support and system costs as highlighted by the superiority of the solution obtained with A1 and A2 as compared to A0.

\subsection{Scalability}
\begin{figure}[!b]
\vspace{-2.0em}
    \centering
    \includegraphics[width=0.50\textwidth]{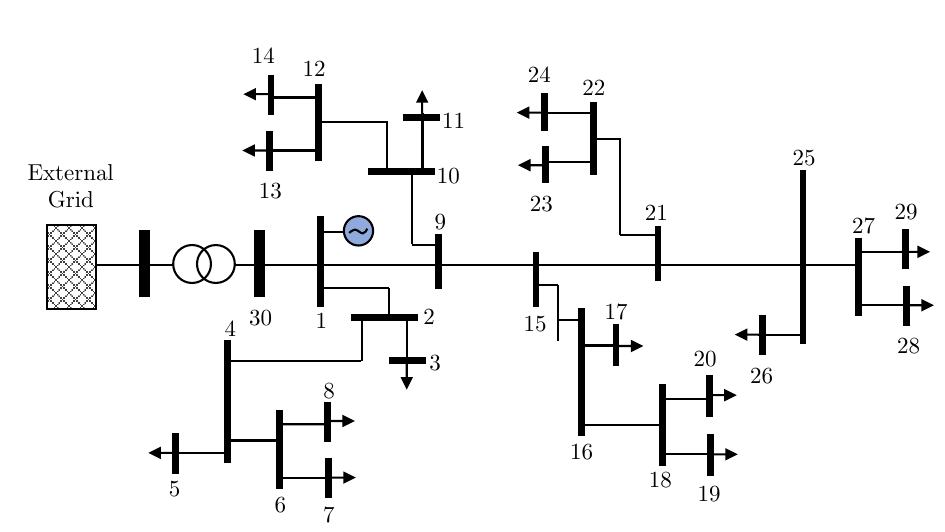}
    \caption{Medium voltage 30-bus test network.}
    \label{fig:32_bus}
\end{figure}
The algorithm is further tested on a 30-bus MV distribution network shown in Fig.~\ref{fig:32_bus}, with the network parameters described in~\cite{epfl_th}. The 30-bus network consists of one pre-installed SG unit at node one. Furthermore, seven units are considered as investment candidates at nodes =\{3,\,6,\,12,\,18,\,24,\,27\}, consisting of one SG unit, two VSM-based CIG units (PV$_1$), two droop-based CIG units (PV$_2$), and two fixed power output CIG units (PV$_3$). The performance of the different algorithms is presented in Table \ref{tab:30_bus} considering four representative days. 

The results indicated in Table \ref{tab:30_bus} are obtained for the optimal solution of the 30-bus network with varying RoCoF limits. Both A1 and A2 outperform A0 in terms of total planning costs and total computation time. It is observed that while A1 and A2 still indicate identical investment solutions, the former provides lower total costs in each case. However, in both cases, A2 is observed to be more computationally efficient as compared to A1. The choice between A1 and A2 therefore will depend on the choice between computational efficiency and optimality of the final solution.

\begin{table}[t!]
 \renewcommand{\arraystretch}{1.2}
 \caption{Planning solutions for the 30-bus network with varying RoCoF limits.}
    \begin{subtable}[h]{1.00\linewidth}
        \centering
        \caption{Case A: RoCoF limit = 2 Hz/s}
        \resizebox{0.9\linewidth}{!}{
       \begin{tabular}{l|ccc|}
\cline{2-4}
 & \multicolumn{1}{c|}{\textbf{Algorithm 0}} & \multicolumn{1}{c|}{\textbf{Algorithm 1}} & \textbf{Algorithm 2} \\ \toprule \hline
 \multicolumn{4}{|c|}{Costs and decisions} \\ \hline
\multicolumn{1}{|l|}{Total cost (\$)} & \multicolumn{1}{c|}{3045089} & \multicolumn{1}{c|}{2881913} & 2960361 \\ \hline
\multicolumn{1}{|l|}{Investment decision} & \multicolumn{1}{c|}{$2\times$PV$_3$, $2\times$SG} & \multicolumn{1}{c|}{$2\times$PV$_2$, $2\times$SG} & $2\times$PV$_2$, $2\times$SG \\ \hline\bottomrule
 \multicolumn{4}{|c|}{Frequency support} \\ \hline
\multicolumn{1}{|l|}{M (s)} & \multicolumn{1}{c|}{15.60} & \multicolumn{1}{c|}{15.60} & 15.60 \\ \hline
\multicolumn{1}{|l|}{D (p.u)} & \multicolumn{1}{c|}{1.17} & \multicolumn{1}{c|}{34.50} & 34.50 \\ \hline\bottomrule
 \multicolumn{4}{|c|}{Computational performance} \\ \hline
\multicolumn{1}{|l|}{Computation time (s)} & \multicolumn{1}{c|}{ 7255} & \multicolumn{1}{c|}{4599} & 4377 \\ \hline
\end{tabular}}
       \label{tab:roco30-1}
    \end{subtable}
    \vfill \vspace{1.5em}
    \begin{subtable}[h]{1.00\linewidth}
        \centering
        \caption{Case B: RoCoF limit = 1 Hz/s}
        \resizebox{0.9\linewidth}{!}{
        \begin{tabular}{l|ccc|}
\cline{2-4}
 & \multicolumn{1}{c|}{\textbf{Algorithm 0}} & \multicolumn{1}{c|}{\textbf{Algorithm 1}} & \textbf{Algorithm 2} \\ \toprule\hline
 \multicolumn{4}{|c|}{Costs and decisions} \\ \hline
\multicolumn{1}{|l|}{Total cost (\$)} & \multicolumn{1}{c|}{3760089} & \multicolumn{1}{c|}{3131387} & 3140962 \\ \hline
\multicolumn{1}{|l|}{Investment decision} & \multicolumn{1}{c|}{\begin{tabular}[c]{@{}c@{}}$2\times$PV$_3$, $1\times$PV$_3$,\\ $2\times$SG\end{tabular}} & \multicolumn{1}{c|}{\begin{tabular}[c]{@{}c@{}}$1\times$PV$_1$, $1\times$PV$_3$,\\ $2\times$SG\end{tabular}} & {\begin{tabular}[c]{@{}c@{}}$1\times$PV$_1$, $1\times$PV$_3$,\\ $2\times$SG\end{tabular}} \\ \hline \bottomrule
 \multicolumn{4}{|c|}{Frequency support} \\ \hline
\multicolumn{1}{|l|}{M (s)} & \multicolumn{1}{c|}{15.60} & \multicolumn{1}{c|}{22.60} & 22.60 \\ \hline
\multicolumn{1}{|l|}{D (p.u)} & \multicolumn{1}{c|}{17.84} & \multicolumn{1}{c|}{1.47} & 1.47 \\ \hline\bottomrule
 \multicolumn{4}{|c|}{Computational performance} \\ \hline
\multicolumn{1}{|l|}{Computation time (s)} & \multicolumn{1}{c|}{6745} & \multicolumn{1}{c|}{5048} & 4585 \\ \hline
\end{tabular}}
        \label{tab:roco30-2}
     \end{subtable}
     
     \label{tab:30_bus}
\end{table}

\section{Conclusion}\label{concl}
The abrupt disconnection of MGs from the main grid can trigger cascading failures caused by the activation of different protective devices in the network. It is therefore vital that the effect of these high-impact events is embedded in the planning model to ensure system security and resilience. However, most of the existing investment planning methods only consider the static pre- and post-fault security constraints while ignoring the transient security of the MG during the islanding event. In this paper, two tractable algorithms are proposed to provide a cost-effective transiently secure solution incorporating the frequency behavior after islanding into the planning problem. The proposed algorithms adopt effective and efficient decomposition strategies to integrate non-linear transient frequency security constraints tractably. Moreover, the solution algorithms proposed concurrently optimize the required level of frequency support and system costs. 

The model presented however utilizes the CoI approximation ignoring the local frequency response of the different generators. Therefore, future works will modify the model to include the effect of local frequency response on planning decisions. In addition, variations in control parameters due to the intermittency of renewable-powered CIG units need to be adequately captured to further enhance the robustness of the model.

%
\vspace{-0.25cm}
\bibliographystyle{IEEEtran}

\end{document}



\begin{tikzpicture}[auto,line/.style ={draw, >=latex'}]

  \begin{groupplot}[
    group style={
    group name=my plots,
    group size=1 by 2,
    vertical sep=10pt,
    },
    footnotesize,
    scaled ticks=false,
    tick label style={/pgf/number format/fixed},
    width=7cm,
    height=3.5cm,
    enlarge x limits=0.1,
    ]
    
    \nextgroupplot[
    ybar,
    enlarge x limits=0.025,
	bar width=10pt,
	ymin=0, ymax=40,
	xmin=0.5, xmax=3.5,
    legend style={at={(0.485,1.105)},anchor=south,legend columns=3},
    legend cell align={left},
    xtick={1,2,3},
    xticklabel = \empty,
    ylabel={\footnotesize  Inertia [s]},
    ]
    \footnotesize

    \addplot+[ybar, color=black, fill=pYellow] plot coordinates {(1,7.84) (2,17.64) (3,17.64)};
    \addplot+[ybar, color=black, fill=pGreen] plot coordinates {(1,22.28 ) (2,34.66) (3,34.66)  };
     \addplot+[ybar, color=black, fill=pPurple] plot coordinates {(1,22.28) (2,22.28) (3,22.28) };

    \legend{{\scriptsize Case A$\;\;$},{\scriptsize Case B$\;\;$},{\scriptsize Case C$\;\;$}};
    
\nextgroupplot[
    enlarge x limits=0.025,
    ybar ,
	bar width=10pt,
	ymin=0, ymax=30,
    legend style={at={(0.4,1.075)},anchor=south,legend columns=2},
    legend cell align={left},
    xmin=0.5, xmax=3.5,
    xtick={1,2,3},
    xticklabels={A0,A1, A2},
    ylabel={\footnotesize  Damping [p.u]},
    ]
    \footnotesize

     \addplot+[ybar, color=black, fill=pYellow] plot coordinates {(1,18.00) (2,1.13 ) (3,1.13 )};
    \addplot+[ybar, color=black, fill=pGreen] plot coordinates {(1,23.54  ) (2,2.23) (3,2.23)  };
     \addplot+[ybar, color=black, fill=pPurple] plot coordinates {(1,23.54 ) (2,23.54 ) (3,23.54 ) };

  \end{groupplot}

\end{tikzpicture}